\documentclass{emulateapj}
\usepackage{mathptmx}

\slugcomment{31 October 2005}

\shorttitle{Characteristics of Doppler Shift Oscillations}
\shortauthors{Mariska}

\newcommand{\Caxix}{\ion{Ca}{19}}
\newcommand{\Sxv}{\ion{S}{15}}

\begin{document}

\title{Characteristics of Solar Flare Doppler Shift Oscillations
  Observed with the Bragg Crystal Spectrometer on \textit{Yohkoh}}
\author{John T. Mariska}
\affil{E.\ O.\ Hulburt Center for Space Research, Code 7673,
Naval Research Laboratory, Washington, DC 20375}
\email{mariska@nrl.navy.mil}

\begin{abstract}
  This paper reports the results of a survey of Doppler shift
  oscillations measured during solar flares in emission lines of
  \ion{S}{15} and \ion{Ca}{19} with the Bragg Crystal
  Spectrometer (BCS) on \textit{Yohkoh}.  Data from 20 flares
  that show oscillatory behavior in the measured Doppler shifts
  have been fitted to determine the properties of the
  oscillations.  Results from both BCS channels show average
  oscillation periods of $5.5 \pm 2.7$ minutes, decay times of
  $5.0 \pm 2.5$ minutes, amplitudes of $17.1 \pm 17.0$ km
  s$^{-1}$, and inferred displacements of $1070 \pm 1710$ km,
  where the listed errors are the standard deviations of the
  sample means.  For some of the flares, intensity fluctuations
  are also observed.  These lag the Doppler shift oscillations by
  $1/4$ period, strongly suggesting that the oscillations are
  standing slow mode waves.  The relationship between the
  oscillation period and the decay time is consistent with
  conductive damping of the oscillations.
\end{abstract}


\section{INTRODUCTION}

The solar corona is a high-temperature, fully-ionized plasma
structured by magnetic forces.  This structuring results in the
familiar loop-dominated appearance of the corona.  The fact that
the corona is threaded by a magnetic field also means that it is
subject to a rich assortment of oscillatory modes, which have
been the subject of considerable theoretical investigation
\citep[e.g.,][]{Roberts1983,Roberts1984}.  \citet{Roberts2000},
and \citet{Roberts2003} provide recent reviews of the theoretical
state of the subject. 

Searches for coronal oscillations have a long history, with much
of the earliest work centered on looking for evidence of the
propagation of the waves that were thought to heat the corona
\citep[for a review see, e.g.,][]{Aschwanden2003}.  But it has
been the more recent observations from spacecraft imaging
instruments, especially the \textit{Transition Region and Coronal
  Explorer} (\textit{TRACE}) that have begun to expose the rich
variety of oscillatory phenomena present in the corona.
\textit{TRACE} observations of spatial shifts of 1~MK loops in a
flaring active region showed unambiguously that periodic
oscillations were occurring, which \citet{Aschwanden1999}
identified as MHD kink mode oscillations.

Spectroscopic instruments have also detected coronal
oscillations.  Observations obtained with the Solar Ultraviolet
Measurements of Emitted Radiation (SUMER) spectrometer on
\textit{SOHO}, show the presence of damped oscillatory Doppler
shifts in the emission lines of \ion{Fe}{19} at 1118.1~\AA\ and
\ion{Fe}{21} at 1354.1~\AA\ \citep{Wang2002,Kliem2002}.  These
lines are formed at about 8 and 10~MK, respectively.
\citet{Ofman2002} interpreted these observations as evidence for
rapidly-damped slow-mode standing acoustic waves.  Additional
evidence for this conclusion has been provided by
\citet{Wang2003a} and an extensive compilation of SUMER
observations is contained in \citet{Wang2003}.

Both the oscillations observed with \textit{TRACE} and those
observed with SUMER generally damp in a small number of periods,
and there has been considerable speculation about the cause
\citep{Nakariakov1999,Ofman2002b,Ofman2002a,Schrijver2000,Ofman2002}. 
Additional observations, particularly if they cover different
oscillation periods or different solar plasma conditions than
those already measured may provide further constraints on the
damping mechanism, possibly leading to new insights into how the
corona is heated. 

Detection of Doppler shift oscillations with SUMER in emission
lines formed at flare temperatures suggests that other
instruments designed to observe Doppler shifts in flaring plasmas
may be able to detect and characterize these events.  Recently,
\citet{Mariska2005} reported the detection of damped Doppler
shift oscillations in spectra obtained with the Bragg Crystal
Spectrometer (BCS) on \textit{Yohkoh}.  In this paper, I report
on a study of Doppler shift oscillations in a large sample of the
flares observed with that instrument. 

\section{BCS OBSERVATIONS}

The BCS flew on the Japanese \textit{Yohkoh} satellite, taking
useful data from 1991 October 1 to 2001 December 14.  Four bent
crystals viewed the entire Sun over narrow wavelength ranges
centered on emission lines of \ion{Fe}{26}, \ion{Fe}{25}, \Caxix,
and \Sxv, with the two Fe bands sharing one detector and the Ca
and S bands sharing a second.  Useful data were rarely recorded
in the \ion{Fe}{26} band, and the spectrum observed in the
\ion{Fe}{25} band is complex---consisting of contributions from
several stages of ionization of Fe.  Thus, this study was
restricted to data from the \Caxix\ and \Sxv\ wavelength bands. 
Further details on the characteristics of the BCS are provided by
\citet{Culhane1991}. 

To search for Doppler shift oscillations, plots of the total
count rate in the BCS \Caxix\ channel as a function of time for
the first year of operations, 1991 October 1 through 1992
September 30, were scanned for candidate flares.  Any flare with
peak count rate in the \Caxix\ channel of more than about 1000
counts s$^{-1}$ and a smooth profile suggesting that the event
was not seriously contaminated by emission from other events on
the solar disk was processed with the standard BCS fitting
software and plots of the derived physical parameters as a
function of time were examined visually for evidence of
oscillatory behavior. 

In addition to flares observed during the first year of
\textit{Yohkoh} observations, all the flares examined in the
study by \citet{Mariska1999} of occulted and nonocculted limb
flares were included.  \textit{TRACE} began routine observations
in 1998 May, and so had some overlap in time with the
\textit{Yohkoh} mission.  All the flares listed in
\citet{Schrijver2002} and \citet{Aschwanden2002} that were also
observed with the BCS were therefore also included in this
study.  The \textit{Yohkoh} mission also overlapped in time with
the \textit{SOHO} mission.  A search of the BCS data, however,
shows no useful observations for any of the events observed by
\citet{Wang2003}. 

\begin{figure}
\plotone{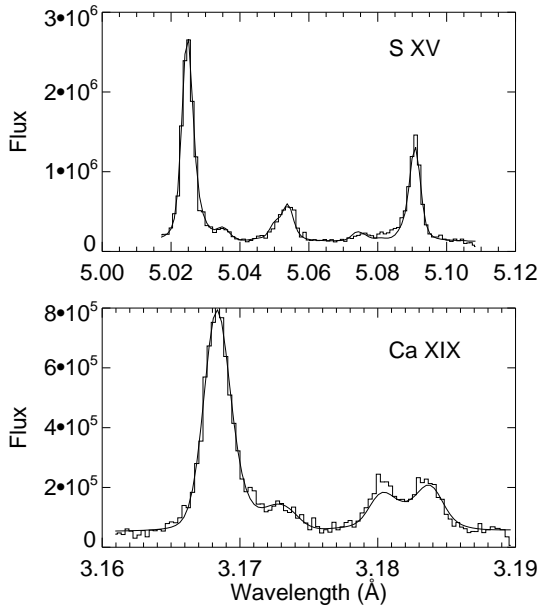}
\caption{BCS spectra in the \ion{S}{15} and \ion{Ca}{19} channels
  obtained on 1991 October 21 at 2:48 UT.  Fluxes are in units of
  photons cm$^{-2}$ s$^{-1}$ \AA$^{-1}$.  The data are plotted as
  histograms and the best-fit model as a smooth line.  Note that
  the wavelength axis for both plots is based on the nominal
  \textit{Yohkoh} pointing and assumes the flare is near the
  equator.  Thus the absolute values of wavelength are not
  precise.} 
\label{fig:bcs_spectra}
\end{figure}

Figure~\ref{fig:bcs_spectra} shows examples of typical spectra
obtained in the \Sxv\ and \Caxix\ channels for a flare observed
at roughly 02:48 UT on 1991 October 21 along with theoretical
spectra fitted to the data.  Although only a few strong features
are obvious in both channels, the spectra are actually quite
complex.  Thus the fitting is done by using detailed atomic
physics data for the transitions that produce lines in the
wavelength ranges of the two channels.  Atomic data for the fits
are primarily from \citet{Bely-Dubau1982a,Bely-Dubau1982b} and
\citet{Vainshtein1978,Vainshtein1985}.  The required ionization
equilibrium calculations are from \citet{Arnaud1985}. 

The \Sxv\ and \Caxix\ resonance lines, which are the strongest
lines in each channel, are formed over relatively broad
temperature ranges.  The contribution functions peak at about
15.8 and 31.6~MK, respectively.  In typical flares, however, most
of the emission comes from plasma with temperatures less than
about 20~MK, so the emission observed in the BCS \Caxix\ band
tends to be from plasma at lower temperatures than the peak of
the contribution function.  The key feature of the emission in
both bands, is that there are line ratios in each band that are
temperature sensitive.  Thus, carefully fitting all the emission
lines in each band provides an excellent measurement of the
temperature of the emitting plasma.

Each spectrum is fitted using an isothermal model and the
standard BCS fitting software.  Thus the fit is characterized by
a temperature $T$, an emission measure EM, a nonthermal
broadening $\xi$, and a Doppler shift velocity $v$.  The best-fit
values for the parameters are determined using
Levenberg-Marquardt least-squares minimization
\citep[e.g.,][]{Bevington1969}.  For the \Sxv\ spectrum shown in
the figure.  the best fit parameters for $T$, EM, and $\xi$ are
12.6~MK, $2.30\times10^{48}$~cm$^{-3}$, and 38.1 km s$^{-1}$,
respectively.  For the \Caxix\ spectrum, the corresponding values
are 13.6~MK, $2.30\times10^{48}$~cm$^{-3}$, and 76.7 km s$^{-1}$. 
As Figure~\ref{fig:bcs_spectra} shows, the fits are generally
excellent.  The different values for $T$ and $\xi$ suggest,
however, that the two wavelength channels are not sampling
exactly the same plasma. 

The BCS is uncollimated and therefore views the entire Sun.  When
only one flare is taking place, this does not present any
problems, since the high-temperature plasma is generally
concentrated in a small enough volume that very little line
broadening due to the spatial extent of the plasma is present. 
Because the angle of incidence of the emission into the
spectrometer depends on the flare's location, the reference
wavelength for each flare will be different.  The wavelength
scale shown on the spectra in Figure~\ref{fig:bcs_spectra} is for
the nominal spacecraft pointing and a flare near the equator. 
Since this study is concerned with Doppler shift variations
within each individual flare, the lack of an absolute reference
wavelength is unimportant. 

In its normal mode of operation, the BCS obtains spectra in all
four wavelength channels every 3~s.  Early or late in flares,
when the count rates can be quite low, the individual spectra are
difficult to fit.  For this work, I have therefore summed
individual spectra until a minimum of 10,000 counts are
accumulated in each channel.  This generally results in excellent
fits to the data and assures that each data point in the
resulting time series will have roughly the same statistical
significance.  Note that these accumulated spectra are produced
for each channel individually.  Thus the time series of fitted
spectra for the two channels will not generally be at identical
times. 

BCS data for a total of 103 flares were processed in the manner
described above and examined for evidence of oscillatory
behavior.  Of this total, 38 showed evidence of oscillatory
behavior in the measured Doppler shifts that looked promising
enough to analyze further.  Of those 38 flares, only 20 were
deemed suitable for inclusion in this paper.  While the remaining
18 flares could be fitted using the functional form developed in
the next section, the resulting fits were less convincing than
the 20 included flares.  Generally, this was due to the
Doppler-shift data exhibiting less than a complete oscillation
period, leading to considerable ambiguity in the decay time
determination. 

\section{RESULTS}

\begin{figure*}
\plotone{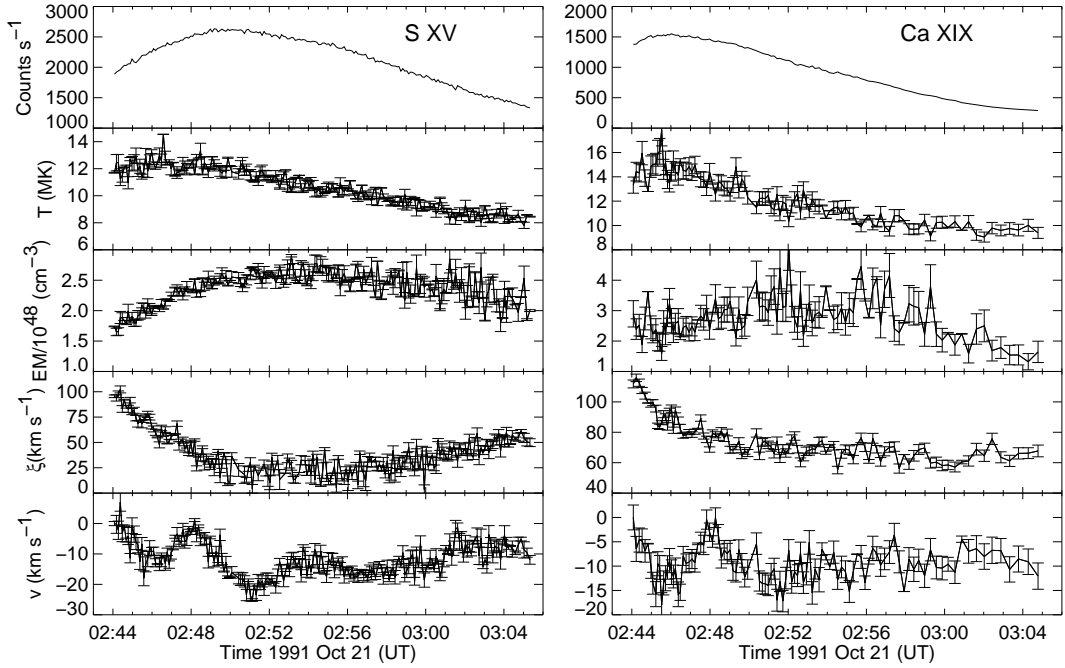}
\caption{Temporal behavior of the intensity, temperature,
  emission measure, nonthermal broadening velocity, and Doppler
  shift derived from the BCS \ion{S}{15} and \ion{Ca}{19}
  observations for the 1991 October 21 flare.  The zero value for
  the Doppler shift velocities has been set to the wavelength
  shift of the first fitted spectrum.} 
\label{fig:bcs_data}
\end{figure*}

Figure~\ref{fig:bcs_data} shows the temporal behavior of the
count rates and the fitted parameters from the fits to the
accumulated spectra for the \Sxv\ and \Caxix\ wavelength channels
for one of the flares that exhibited damped oscillatory behavior. 
The count rates, temperatures, emission measures, and nonthermal
velocities show relatively smooth behavior.  In both channels,
however, the Doppler shifts show clear evidence for oscillatory
behavior.  In particular, the Doppler shift as a function of time
in the \Sxv\ channel appears to show clearly evidence for a
damped oscillation. 

Both the BCS \Sxv\ and \Caxix\ wavelength channels are within the
\textit{GOES} 1--8~\AA\ band, and the \Sxv\ intensity curve shown
at the top left of Figure~\ref{fig:bcs_data} generally tracks the
intensity measured with \textit{GOES} in that band.  For example,
the peak flux in the \textit{GOES} 1--8~\AA\ band occurred at
2:50~UT, the same time as the peak in the BCS \Sxv\ band.  The
emission in the BCS \Caxix\ wavelength channel tends to more
closely track the \textit{GOES} 0.5--4~\AA\ band behavior.  Thus,
the emission in the BCS \Caxix\ wavelength channel peaks earlier
than that seen in the \Sxv\ channel.  As both the \Sxv\ and
\Caxix\ intensity and Doppler-shift data show, oscillatory
behavior is generally present at or before the peak intensity of
the flare and is no longer detectable later in the decay phase.
For flares in which the entire rise-phase is captured, the
oscillations are not initially present, but do develop as the
count rates approach maximum.

Since the BCS is uncollimated, the location on the Sun of each
flare determines where on the detectors the spectrum falls and
there is no absolute wavelength reference.  In the past, most
analyses of Bragg crystal spectrometer data have adopted the
practice of setting the rest wavelength by assuming that late in
the flare Doppler shifts have disappeared.  Since a cooling
plasma can often result in downflows, this approach is not
without possible ambiguity.  I have therefore chosen to set the
rest wavelength using the first spectrum in each set of flare
observations.  Examination of the velocity plots for the BCS
\Sxv\ and \Caxix\ wavelength channels in
Figure~\ref{fig:bcs_data} shows that for this flare, assuming
that the zero for the velocity scale can be set using a
measurement late in the flare, suggests that early in the event
the measured velocities represent roughly 10~km~s$^{-1}$ upflows.
The relationship between these small Doppler shifts and the much
larger Doppler shifted secondary component that appears very
early in many flares is not clear.  Thus considerable caution
should be exercised in drawing any connection between the overall
trend in the Doppler shifts and a possible driver for the
oscillations.

Following the example of the earlier work on spatial
oscillations observer with \textit{TRACE}
\citep[e.g.,][]{Aschwanden2002}, I fit the BCS Doppler shift
observations with a combination of a damped sine wave and a
polynomial background.  Thus, for each channel, I assume that the
data can be fit with a function of the form
\begin{equation}
v(t) = A_0 \sin(\omega t + \phi)\exp(-\lambda t) + B(t)\, ,
\label{eq:sine_wave}
\end{equation}
where
\begin{equation}
B(t) = b_0 + b_1 t + b_2 t^2 + b_3 t^3 + \cdots
\label{eq:background}
\end{equation}
is the trend in the background data.  Beginning with an initial
guess for the fit parameters, the data in each channel were
fitted to equation~(\ref{eq:sine_wave}) using Levenberg-Marquardt
least-squares minimization \citep[e.g.,][]{Bevington1969}.  For
all the flares studied the number of terms needed for the
background expression was generally one or two, although
occasionally three terms produced a better fit. 

\begin{figure*}
\plotone{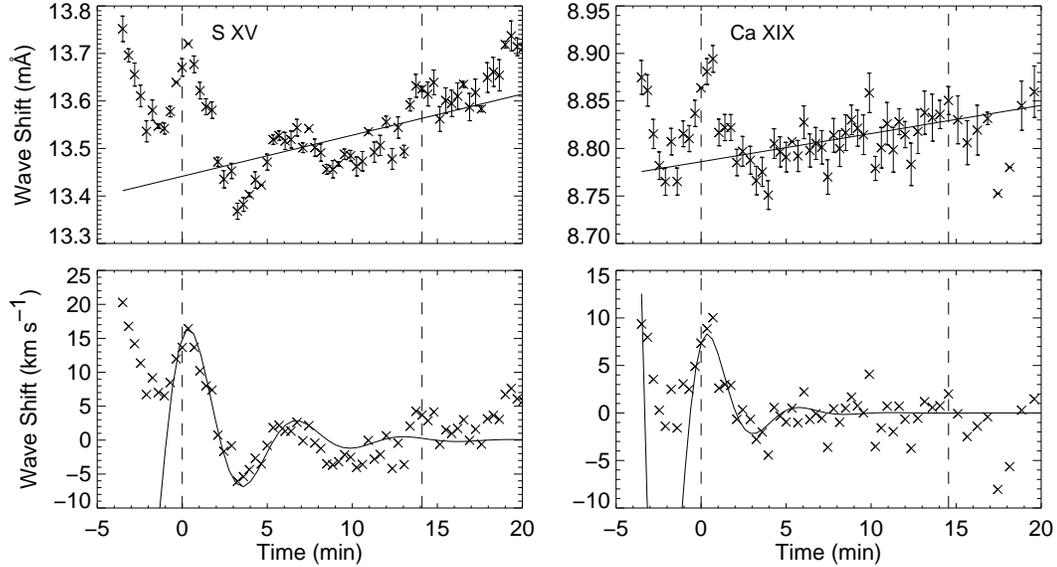}
\caption{Averaged BCS Doppler shift data and decaying sine wave
  fits for the \ion{S}{15} and \ion{Ca}{19} observations for the
  1991 October 21 flare.  The zero value for the shifts plotted
  in the top panels is based on an arbitrary reference system for
  the BCS in which a flare near the solar Equator with no Doppler
  shift would have a value of 0~m\AA.  Time is measured from
  02:47:34 UT.} 
\label{fig:bcs_fits}
\end{figure*}

While the \Sxv\ Doppler shift data were generally quite smooth,
often the \Caxix\ data showed considerable fluctuations on top of
the apparent damped oscillation.  These fluctuations are present
even though the data have been accumulated to more than 10,000
total counts in the channel, and are probably a manifestation of
the fact that not all the high-temperature flare plasma is taking
part in the oscillation.  In some cases these secondary
fluctuations were sufficiently large that it was difficult to fit
the time series using equation~(\ref{eq:sine_wave}).  For the
flares that showed obvious oscillatory behavior, the periods
appear to be on the order of a few minutes.  Thus, to improve the
fits, I have accumulated the BCS data for longer intervals by
requiring that the minimum number of counts in a channel exceed
10,000 and that the accumulation time be at least 20~s.  Note
that since for most BCS data sets the minimum accumulation time
is 3~s, this means that the data analyzed in this study are
generally accumulated for at least 21~s.  The top panels in
Figure~\ref{fig:bcs_fits} show the Doppler shift data for the
\Sxv\ and \Caxix\ channels for the flare data shown in
Figure~\ref{fig:bcs_data} accumulated in this manner along with
the best-fit background trend.  For both channels, the best-fit
background was a two-term polynomial. 

The background-subtracted data along with the best-fit damped
sine wave are shown in the bottom two panels of
Figure~\ref{fig:bcs_fits}.  In each wavelength channel, only the
regions between the vertical dashed lines were fitted.  While
that data accumulated for longer times are still noisy, the fits
appear to be reasonable.  For the \Sxv\ and \Caxix\ channels the
reduced $\chi^2$ values are 5.8 and 1.2, respectively. 

Table~\ref{table:properties} summarizes the results of fitting
all the flares examined in this study for which reasonable fits
to a damped sine wave were possible.  For each flare, the table
lists the date and time of the first \Sxv\ spectrum that was
included in the fit, the class determined from \textit{GOES}
observations, the flare location on the Sun, and the fit
parameters and the $1 \sigma$ error in each parameter.  Flare
locations were taken from the NOAA Space Environment Center daily
reports.  If no position was available from that data source, the
location of the flare was determined using \textit{Yohkoh} Soft
X-Ray Telescope (SXT) data.  Those locations are marked with a
`y' in the table.  For each flare, there are two rows in the
table.  The first row is for the \Sxv\ Doppler shift measurements
and the second row is for the \Caxix\ measurements.  Although the
origin of time for the fits in each channel was tied to the first
fitted spectrum, the values for the \Caxix\ fit parameters in the
table have been adjusted to correspond to the initial time used
for the corresponding \Sxv\ fit.  This makes the fitting
parameters for the two BCS channels directly comparable.  The
errors listed for each parameter in the table are the diagonal
elements in the covariance matrix for the fits. 

Often there are significant Doppler-shift fluctuations before the
time I have selected to begin fitting the data.  The impression
one gets looking at the data is that the initial Doppler-shift
disturbance is somewhat distorted---making it difficult to fit a
decaying sine wave to both this initial fluctuation and the later
more regular signal. 

Examination of the table entries for individual flares shows that
there are events for which all the fit parameters for the two BCS
wavelength channels agree to within the errors and events for
which the fits result in one or more of the fit parameters
differing significantly.  The fit parameters for the 1991 October
21 event shown in Figures~\ref{fig:bcs_data} and
\ref{fig:bcs_fits} are typical of many of the table entries.  The
frequencies and phases for the two channels are close in the two
channels, while the amplitudes and decay rates are somewhat
different, but still close enough to suggest that both channels
are observing the same oscillation.  For some events, though,
most of the fit parameters differ significantly, suggesting that
the two BCS wavelength channels are sampling different
oscillations.  The 1993 September 26 flare is an example. 

\begin{figure*}
\plotone{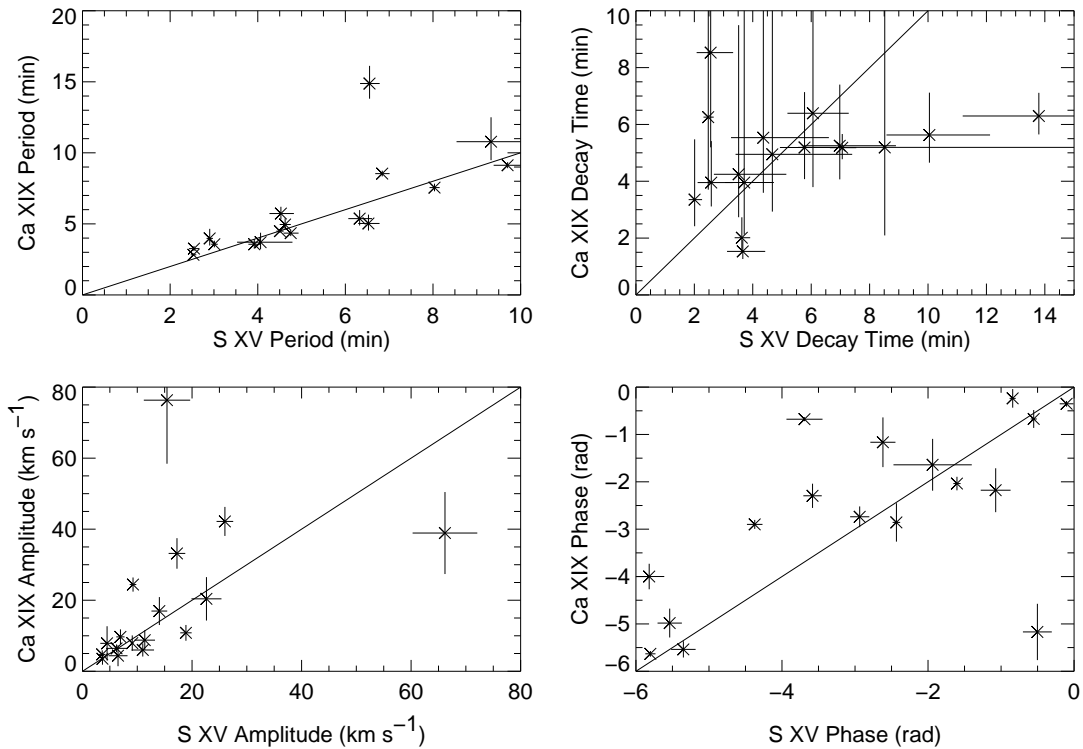}
\caption{Plots of the fitting parameters for events measured in
  the \ion{S}{15} wavelength channel versus the \ion{Ca}{19}
  parameters for the same flare.  Error bars for each point are
  the $1 \sigma$ errors for the fit.  If both channels are
  observing the same event, then the data should scatter around
  the solid lines.} 
\label{fig:s_vs_ca}
\end{figure*}

Since the BCS images the entire Sun and the \Sxv\ and \Caxix\
wavelength channels are sensitive to different temperature
plasmas, it is certainly possible for each channel to be sampling
different structures in the flaring region.  One way to determine
whether this is the case is to produce scatter plots of the fit
parameter in the \Sxv\ channel against the same fit parameter in
the \Caxix\ channel for those flares where data from both
channels could be fitted.  Figure~\ref{fig:s_vs_ca} shows those
plots.  With only a few exceptions, the periods, amplitudes, and
phases, measured in the two BCS wavelength channels suggest that
both channels are sampling plasma from the same structure.  There
is significantly more scatter in the decay time plots, but that
parameter also has large errors associated with it.  The errors
are large for the decay time determination because, as
Figure~\ref{fig:bcs_fits} shows, generally only one cycle or less
of the oscillation is observed.

\begin{figure*}
\plotone{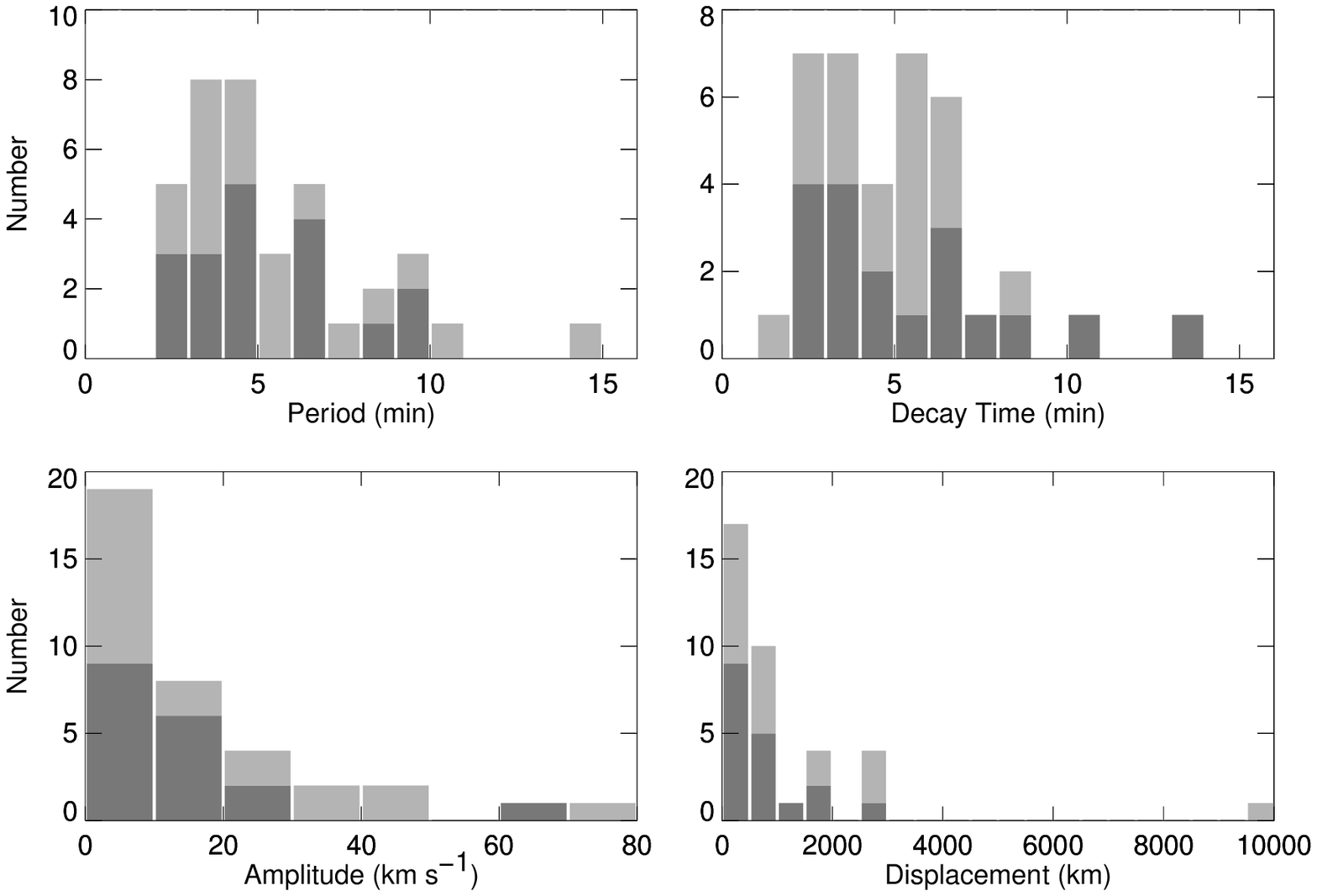}
\caption{Histograms showing the distributions of period, decay
  time, amplitude, and displacement measured from the BCS data. 
  Each bar shows the total number of events in the bin with the
  dark gray showing the \ion{S}{15} contribution and the light
  gray showing the \ion{Ca}{19} contribution.} 
\label{fig:histograms}
\end{figure*}

Figure~\ref{fig:histograms} shows histograms of the distributions
of the fit parameters and the inferred displacement amplitude for
the events listed in Table~\ref{table:properties}.  Following
\citet{Wang2003}, we define the maximum displacement amplitude
for each fit as
\begin{equation}
A = A_0/(\omega^2 + \lambda^2)^{1/2} \, . 
\end{equation}
The distributions shown in the figure appear to be similar for
both BCS wavelength channels.  Table~\ref{table:average_values}
lists average values and standard deviations for the parameters
in each channel and the combined data.  Also listed in the table
are the average properties for oscillations observed with
\textit{TRACE} from \citet{Aschwanden2002} and observed with
SUMER from \citet{Wang2003}.  A reanalysis of the SUMER data
gives similar values to those in the table \citep{Wang2005}. 

\section{BCS OSCILLATION MODES}

The transverse loop oscillations observed with \textit{TRACE} are
thought to be fast kink mode MHD oscillations
\citep{Aschwanden2002}.  Those observed with SUMER have been
interpreted as slow standing mode MHD waves \citep{Wang2003a}. 
Comparison of the averages in Table~\ref{table:average_values}
with the \textit{TRACE} and SUMER results suggests that the BCS
values are closer to those from \textit{TRACE} than those from
SUMER.  The overall picture, however, is not entirely clear. 
Because exposure times for the SUMER observations tended to be
quite long ($>50$~s), those observations are not sensitive to the
shortest periods observed with the BCS.  On the other hand, as
shown in Figure~\ref{fig:bcs_data}, the BCS data often only
extend for 20~minutes or less.  Thus, for many events, they would
not be sensitive to the longer periods seen in the SUMER data. 
It might be speculated, though, that the slow Doppler shift
evolution seen in the \Sxv\ data in Figure~\ref{fig:bcs_data} and
partially removed by the background polynomial fits to the data
is actually a portion of a longer period damped oscillation that
is superimposed on those being measured in this paper. 

For a loop of length $L$, the period of a slow-mode wave is given
by
\begin{equation}
P = \frac{2L}{j c_T} \, ,
\label{eq:standing_mode}
\end{equation}
where $j$ is the node number, usually assumed to be 1, and $c_T$
is the slow magnetoacoustic speed, which is close to the sound
speed \citep{Roberts1984}.  Early in the time period where the
BCS Doppler shift oscillations are observed, the temperatures
measured in the \Sxv\ channel are around 12~MK, while those
measured in the \Caxix\ channel are around 14~MK (see, e.g.,
Figure~\ref{fig:bcs_data}).  For a coronal composition plasma,
the sound speed is given by $c_s \approx 0.152 \sqrt T$ km
s$^{-1}$.  Thus, if the oscillations are slow standing mode
waves, the average periods in the table and their standard
deviations imply loop lengths of about $82 \pm 35$~Mm for \Sxv\
and $99 \pm 55$~Mm for \Caxix---roughly 113 and 137 arcsec,
respectively.  For semicircular loops with the above lengths,
this corresponds to loop radii of 26 and 32~Mm, respectively, or
36 and 44 arcsec.

The plasma observed with the BCS \Sxv\ and \Caxix\ channels tends
to have higher temperatures than those measured for the same
events using the \textit{Yohkoh} soft X-ray telescope (SXT)
\citep[e.g.,][]{Doschek1999}.  Thus, it is not clear that the the
sizes of the loops observed with SXT are appropriate for
comparison with the above estimates.  Moreover, many of the
flares are at the limb and have portions of the loop occulted by
the solar disk. 

There are, however, some interesting events.  The 1992 July 17
flare was not occulted and clearly showed a loop-like structure.
This flare was included in a study of occulted and nonocculted
flares by \citet{Mariska1999}.  The loop is near the limb, so
projection effects make a length determination difficult.  A.
Winebarger (2005, private communication) has used the SXT Be
filter data for this flare to examine possible geometries for the
loop.  The geometry is difficult to define, but if the aspect
ratio is near one (a semicircular loop), the length would be
approximately 60~Mm.  The frequencies listed for this flare in
Table~\ref{table:properties} imply lengths of 47 and 61~Mm for
the \Sxv\ and \Caxix\ observations, respectively.  Given the
uncertainties in connecting the SXT images to the plasma being
observed with the BCS \Sxv\ and \Caxix\ channels, these numbers
are in excellent agreement.

If the oscillations are slow standing mode waves, other
observable physical quantities should also be oscillating.  For a
sound wave, $v = c_s \delta\rho/\rho$.  Thus, for a 12~MK plasma,
$\delta\rho/\rho \approx 0.04$, a very small density fluctuation.
Even though the intensity is proportional to the square of the
density, this is a small fluctuation.  Also, since the BCS images
the entire loop, it would be difficult to see this small an
oscillation when the entire loop is observed since for the
fundamental mode the density oscillation has anti-nodes at the
two footpoints of the loop, and the oscillations at these two
anti-nodes are in anti-phase.  For the partially occulted flares,
one might expect, however, to see intensity fluctuations.

\begin{figure*}
\plotone{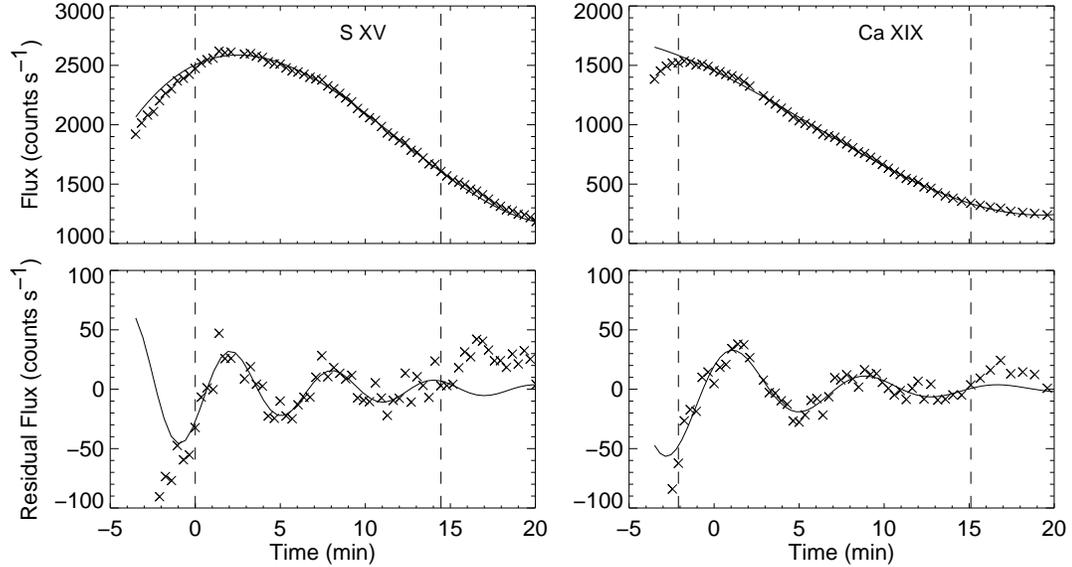}
\caption{Averaged BCS \ion{S}{15} and \ion{Ca}{19} intensity data
  and decaying sine wave fits for the 1991 October 21 flare. 
  Only the data between the vertical dashed lines have been
  fitted.  Time is measured from 02:47:34 UT---the same starting
  time used for Figure~\ref{fig:bcs_fits}.} 
\label{fig:bcs_intens_fits}
\end{figure*}

The four panels of Figure~\ref{fig:bcs_intens_fits} show the
results of fitting a polynomial background and exponentially
decaying sine wave to the intensity data for the 1991 October 21
flare.  Only the data between the vertical dashed lines were
fitted.  The top panels show the intensity data and the best-fit
background trend.  In both BCS channels, the background was fit
with a cubic polynomial.  Error bars are not included in the
plots.  Since the integration times for the portion of the data
used in the fits were generally 21~s, the errors for a count rate
of 2500 counts s$^{-1}$ are about 11 counts s$^{-1}$.  For a 1500
counts s$^{-1}$ count rate, the corresponding error is about 8
counts s$^{-1}$.  Thus the fluctuations fitted in the bottom two
panels are well above the level expected from counting
statistics. 

Table~\ref{tab:intens_results} summarizes the intensity fitting
results for this flare.  While the origin of the time values for
the fitting calculation in each channel was tied to the time of
the first fitted spectrum, the results in the table for the
\Caxix\ channel have been adjusted to correspond to the initial
time used for the \Sxv\ fits.  The beginning time for the \Sxv\
intensity fits is the same as the beginning time used for the
velocity fits for this flare listed in
Table~\ref{table:properties}.  Thus the intensity results for the
two channels can be directly compared with each other and with
the results for the same flare shown in
Table~\ref{table:properties}. 

The periods and decay times of the intensity fluctuations are
comparable with the characteristics of the fits to the Doppler
shift measurements for this flare.  The phases for the fits to
the intensity measurements are, however, different.  This is most
easily visualized by comparing the times of the first peak in the
fits to the Doppler shifts in Figure~\ref{fig:bcs_fits} with
those of the fits to the intensity fits in
Figure~\ref{fig:bcs_intens_fits}.  For \Sxv, the intensity peak
occurs 1.54 minutes after the Doppler shift peak.  This
corresponds to a delay of 1/4 of the period, strongly suggesting
that the oscillations observed in this flare are due to slow
standing-mode MHD waves \citep{Sakurai2002}. 

Additional evidence for intensity oscillations in the 1991
October 21 flare is provided by data from the \textit{GOES}
satellite.  One-minute integration time data for this flare are
available from the National Geophysical Data Center.  Polynomial
fits to the same time intervals used for the \Sxv\ intensity data
for this flare reveal the same oscillations in the residual
signal in both the 1--8 and 0.5--4~\AA\ \textit{GOES} channels. 

One curious feature of the intensity fluctuations appears to be a
tendency for the oscillation in the residual signal to be more
obvious in the data from higher temperature plasma.  Note, for
example in Figure~\ref{fig:bcs_intens_fits}, that the
oscillations in the \Caxix\ intensity are better defined than
those in the \Sxv\ residuals.  This also appears to be the case
for the two \textit{GOES} channels.  Examination of the the
intensity data from the BCS \ion{Fe}{25} channel also shows
oscillations in the residual intensities for this flare that are
well defined.  This is probably due to the reduced amount of
background emission at higher temperatures that must be removed
from the oscillatory signal and suggests that oscillations should
be detectable in the lower energy channels for occulted flares
observed with the Hard X-Ray Telescope on \textit{Yohkoh} and
those observed with the \textit{Ramaty High Energy Solar
  Spectroscopic Imager}. 

Intensity fluctuations are not present in all the flares that
exhibit Doppler shift oscillations.  Thus it is worthwhile to ask
whether some of the flares might be exhibiting fast kink mode
oscillations which correspond to the displacement of an entire
flux tube.  For a loop of length $L$, the period of the wave is
given by
\begin{equation}
P = \frac{2L}{j c_k} \, ,
\label{eq:fast_mode}
\end{equation}
where $j$ is the node number, usually assumed to be 1, and $c_k$
is the kink speed \citep{Roberts1984}.  For a flux tube with an
internal magnetic field strength equal to the external field
strength, but a much greater plasma density, $c_k \approx \sqrt 2
c_A$, where $c_A$ is the Alfv\'{e}n speed. 

For the 1992 Jul 17 event discussed above, the estimated loop
length of 60~Mm and a period of 3.3~min (the average of the \Sxv\
channel and \Caxix\ channel results for this event) result in an
Alfv\'{e}n speed of about 600 km s$^{-1}$.  Electron densities
from 10$^{10}$ to 10$^{12}$~cm$^{-3}$ then imply magnetic field
strengths ranging from 27 to 275~G.  For a number of flares
observed with the Naval Research Laboratory S082A spectrograph on
\textit{Skylab}, \citet{Keenan1998} used density sensitive
emission lines of \ion{Ca}{16} to determine an average electron
density of $2.9\times10^{11}$~cm$^{-3}$.  This value would imply
a magnetic field strength of about 128~G.  The low end of this
range of field strengths is consistent with ranges that are
typically quoted for flaring loops \citep[e.g.,][]{Tsuneta1996}. 

Taking a temperature of 14~MK, a density of
$3\times10^{11}$~cm$^{-3}$, and magnetic field strength of 128~G
results in a plasma $\beta$ of about 1.8.  For a density of
$10^{10}$~cm$^{-3}$ and a field strength of 27~G, the plasma
$\beta$ is about 1.3.  The same conclusion was reached by
\citet{Wang2002}, who obtained values near two for $\beta$. 
Generally, the coronal plasma is thought to be low beta plasma. 
Moreover, equation (\ref{eq:fast_mode}) has been derived by
assuming that the plasma is low-$\beta$.  Thus further analysis
of the possibility that some of the observed Doppler shift
oscillations are due to kink modes will require a more careful
examination of the governing equations.  The existence of
intensity fluctuations shifted by $1/4$ phase in some of the
flares argues strongly that all of the flare Doppler shift
oscillations observed with BCS are slow mode standing waves. 

\section{DAMPING OF THE OSCILLATIONS}

Linear slow wave dissipation theory predicts that the decay time
for an oscillation should scale as $P^2$, where $P$ is the
oscillation period \citep{Porter1994}.  More detailed
calculations predict different slopes.  \citet{Ofman2002} have
investigated the damping of the slow-mode oscillations observed
with SUMER using one-dimensional hydrodynamic simulations.  They
considered both damping by thermal conduction and compressive
viscosity and found that for the SUMER observations the damping
was dominated by thermal conduction.  For the conditions in the
loops observed with SUMER, they found that the scaling goes as
$P^{1.17}$ for a loop temperature of 6.3 MK and as $P^{1.07}$ for
a loop temperature of 6.8 MK.  \citet{Ofman2002} attributed the
smaller slope to the nonlinearity of the observed oscillations;
they assumed $A_0/c_s = 0.18$. 

\begin{figure}
\plotone{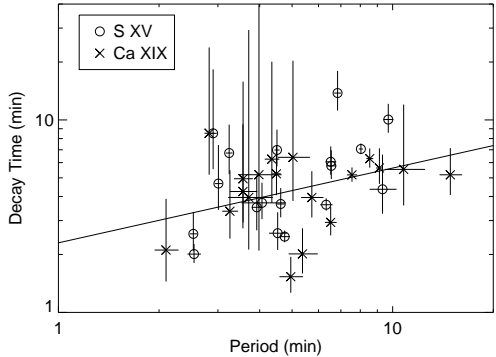}
\caption{The measured decay time vs. period for the events
  observed with the BCS.  Error bars ($1\sigma$) for both the
  period measurement and the decay time measurement are shown. 
  The solid line shows the best fit relationship between the
  log of the period and the log of the decay time.} 
\label{fig:bcs_period_vs_decay}
\end{figure}

Figure~\ref{fig:bcs_period_vs_decay} shows the decay time of the
oscillations plotted against the period for all the events
observed with the BCS.  The line in the plot is the best fit to
all the data and corresponds to $t_{\textrm{decay}} = 2.30
P^{0.39}$---a much smaller slope than that determined by
\citet{Wang2003} from the SUMER observations.  The BCS results,
however, apply to a much higher temperature plasma, 12 to 14 MK,
than the SUMER observations.  (While the peak in the \Sxv\
emissivity function is at about 6.8~MK, the spectral fitting to
the observations, which include temperature-sensitive lines,
results in the higher temperature.)  For the SUMER data, the
highest temperature emission line in which oscillations were
typically observed was from \ion{Fe}{21} with a formation
temperature of about 7.0 MK.  The \citet{Ofman2002} calculations
show that the slope decreases for increasing plasma temperature. 
Since the damping is by conduction, which scales as $T^{2.5}$,
one might expect a sizable change in the slope as the loop
temperature increases.  On the other hand, the oscillations
observed with the BCS have a much smaller amplitude.  For a
temperature of 12 to 14 MK, the average amplitude of 17.1 km
s$^{-1}$ leads to $A_0/c_s \approx 0.04$, somewhat less nonlinear
than is the case for the SUMER oscillations.  Moreover, there is
clearly considerable uncertainty in the BCS results, especially
in the decay time determination. 

\begin{figure}
\plotone{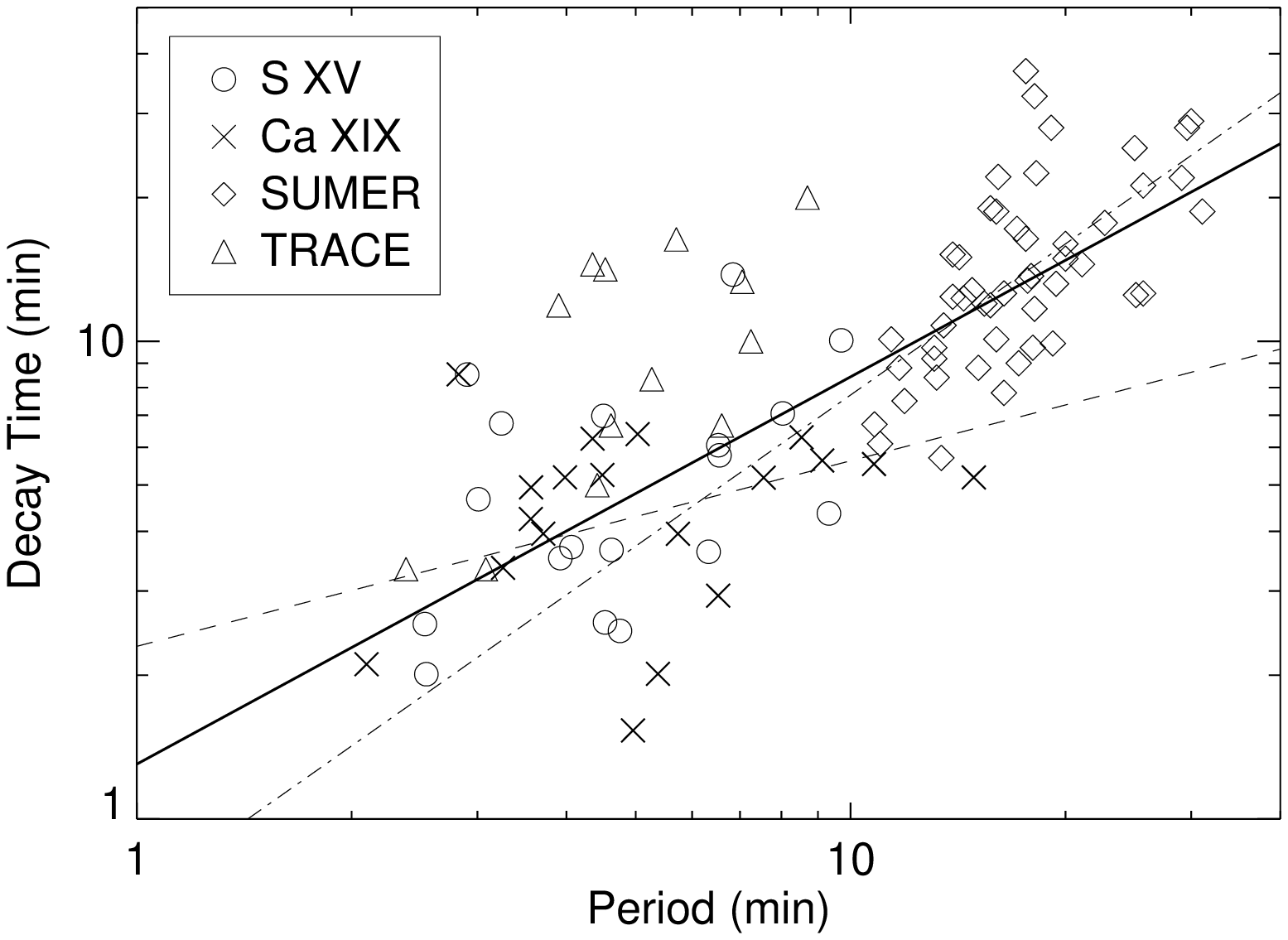}
\caption{The measured decay time vs. period for the events
  observed with the BCS, the SUMER events reported by
  \citet{Wang2003}, and the \textit{TRACE} events reported by
  \citet{Ofman2002a}, \citet{Wang2004}, and
  \citet{Verwichte2004}.  The dashed line shows the best fit
  relationship between the log of the period and the log of the
  decay time plotted in Figure~\ref{fig:bcs_period_vs_decay}, the
  dot-dash line shows a fit to the \citet{Wang2003} data, and the
  solid line shows the fit to the combined BCS and SUMER data
  sets.} 
\label{fig:bcs_sumer_period_vs_decay}
\end{figure}

While the \citet{Ofman2002} calculations suggest that comparisons
between the BCS and SUMER observations may be hindered by the
different formation temperatures of the emission lines being
observed, it is useful to plot all the data together, especially
since the BCS observations explore shorter periods than are
available with the SUMER data. 
Figure~\ref{fig:bcs_sumer_period_vs_decay} shows the decay time
vs. period for the BCS events plotted in
Figure~\ref{fig:bcs_period_vs_decay} along with the results for
events observed with SUMER and \textit{TRACE}.  For the SUMER
data, only the 49 cases fitted by \citet{Wang2003} have been
included.  Note that while the \textit{TRACE} data points lie in
the same period interval on the figure as the BCS data, they
generally have longer decay times than the BCS data---further
suggesting that the \textit{TRACE} data represent a different
kind of oscillation.  Of course, the loops measured in the
\textit{TRACE} observations contain plasma at significantly lower
temperatures, $\sim 1$~MK, than those observed with the BCS and
SUMER. 

Also plotted on the figure are the least squares fits to the BCS
data that are shown on Figure~\ref{fig:bcs_period_vs_decay}, the
SUMER data, and the combined BCS and SUMER data.  The fit to the
combined BCS and SUMER data results corresponds to the expression
$t_{\textrm{decay}} = 1.30 P^{0.81}$. 

\section{DISCUSSION}

It is generally assumed that the emission observed with the
\textit{BCS} is produced in the loop or loops that have been
heated by the flare energy release process.  Thus the observed
oscillations are probably triggered by the heating due to the
flare energy release.  The observation of damped slow-mode
oscillations in the high-temperature flare plasma then presents
interesting challenges for some of the current models for the
evolution of flaring loops.  \textit{TRACE} observations clearly
show that in the decay phase solar flares exhibit many loop-like
structures \citep[e.g.,][]{Warren1999,Warren2000,Aschwanden2001}. 
This has led to attempts to model a flare as a succession of
independently heated threads \citep{Warren2005,Warren2005a}.  To
model a \textit{GOES} M2.0 flare, \citet{Warren2005a} needed 50
individual threads, with each one introduced 50~s after the
previous thread.  Each thread was heated using a triangular
heating function with a width of 200~s. 

The model fits the BCS \Sxv\ and \Caxix\ channels well, yet over
the roughly 5--10 minute period where one sees oscillations in
most of the events discussed here, many individual threads
contribute to the signal observed in the BCS channels.  This
would seem to suggest that either multiple threads contain damped
slow-mode waves with the same parameters, or that an individual
thread is dominating the emission observed over the time interval
where the oscillations are observed.  The model described in
\citet{Warren2005a} is still under development, so it is not
clear, for example, what the time dependent Doppler shifts look
like when the ensemble of threads is considered in their entirety
as might be the case for a flare observed on the disk, or when
only the top portions of the threads are considered as might be
the case for a flare that is partially occulted by the solar
limb.  It is also not yet clear what the minimum number of
threads is that will still smoothly reproduce the light curves
observed with the instruments with which the data have been
compared. 

\textit{Yohkoh} observed the flares that triggered a number of
the events analyzed by \citet{Aschwanden2002}.  Only the 1999
October 25 flare exhibited oscillations in the BCS data that
could be fit using equation~(\ref{eq:sine_wave}).  The three
features observed with \textit{TRACE} exhibited oscillation
periods that ranged from 2.38 to 2.70 minutes, with one feature
showing a decay time of 3.33 minutes.  In contrast, the BCS \Sxv\
and \Caxix\ observations show periods of 9.7 and 9.1 minutes and
decay times of 10.0 and 5.68 minutes, respectively.  Thus, while
the oscillations may be triggered by the same event, it is
unlikely that \textit{TRACE} and the BCS are observing the same
oscillations.  This is not surprising given the significant
differences in temperature sensitivity between the BCS and
\textit{TRACE}. 

\textit{Yohkoh} SXT data are available for 18 of the 20 flares
included in Table~\ref{table:properties}.  Most of the flares are
at the limb and very little looplike structure can be discerned
in the data.  Ten of the flares are included in the
\citet{Sato2003} catalog of flares observed with the
\textit{Yohkoh} SXT and Hard X-Ray Telescope (HXT).  The images
in the catalog made from the HXT data generally show only one
emitting feature---again making it difficult to determine loop
size parameters.  In none of the flares have I been able to
determine whether the oscillations are taking place in a
pre-existing loop or the flaring loop itself.  Some of the SXT
data sets, however, provide extensive, high-cadence observations
of the event.  Studies of the time histories of the intensities
in the individual pixels may provide additional insight into the
nature of the oscillations.  This will be the topic of a future
investigation. 

Of the 103 flares observed by the \textit{Yohkoh} BCS for this
study, only the 20 in Table~\ref{table:properties} exhibited
Doppler-shift oscillations that could be easily fit with a
well-defined decaying sine wave.  As I mentioned earlier, an
additional 18 events could be fitted, but were not used.  Many of
the other events, though, did exhibit significant Doppler-shift
fluctuations.  Since the BCS observes all the flaring plasma at
the temperature of formation of the \Sxv\ and \Caxix\ lines, this
is probably an indication that multiple structures are exhibiting
Doppler shift fluctuations that are out of phase.  While no
attempt was made to observe flares at all center-to-limb
positions on the disk, the results in
Table~\ref{table:properties} clearly suggest the flares near the
limb tend to show well-defined oscillations more frequently than
those closer to disk center.  Since the \textit{Yohkoh} BCS
observes the entire Sun, this may simply imply that flares for
which a portion of the flaring loops are occulted by the solar
limb provide a cleaner Doppler-shift signal. 

Fitting the intensity fluctuations using
equation~(\ref{eq:sine_wave}) is more difficult than fitting the
Doppler-shift fluctuations.  While the background trend in the
Doppler shift data can usually be removed with just one or two
terms in equation~(\ref{eq:background}), the underlying trend in
the intensities usually requires more terms.  If one tries to
detrend the intensity data with the peak in the intensity
included, the background removal is even more complex.  Moreover,
including the background terms as adjustable fitting parameters
often results in the fit selecting a damped sine wave with a
substantial amplitude, up to 25\% of the peak intensity, with a
significantly altered background light curve, which is probably
not a true representation of the oscillation. 

While I will defer a detailed analysis of the intensity
fluctuations to a future paper, a preliminary examination of the
\Sxv\ intensity fluctuations for the 17 flares in
Table~\ref{table:properties} for which \Sxv\ Doppler-shift fits
were available was conducted.  All 17 flares could be fitted with
expressions of the form of equation~(\ref{eq:sine_wave}).  Of
those 17, only six showed phase lags that were judged to be
consistent with that expected for a slow standing-mode MHD wave. 
Fits to five of the remaining flares resulted solutions that were
viewed as having unrealistic background terms.  Fits for the
remaining six flares looked reasonable, but the phases were not
close to the expected 1/4 period lag.  Five of the six flares
with reasonable phase lags were at the limb.  The flares at the
limb generally show only a small concentrated emitting region in
the SXT data. 

Detection of damped intensity fluctuations for some flares in the
\Sxv, \Caxix, and \ion{Fe}{25} wavelength bands and in broad-band
\textit{GOES} observations raises the intriguing possibility that
spatially-resolved coronal loop observations with sufficient
signal-to-noise and short enough exposure times may provide a
wealth of information on the standing slow-mode waves.  It may be
possible to detect damped intensity oscillations in the
\textit{Yohkoh} SXT data.  \citet{McKenzie1997} have searched for
oscillations in SXT image sequences, finding some evidence for
relatively short period fluctuations, 9.6 to 61.6 s.  Their
time-series analysis may, however, have missed the kinds of
intensity fluctuations reported here.  The high sensitivity and
spatial resolution Soft X-Ray Telescope and EUV Imaging
Spectrometer instruments on the upcoming \textit{Solar-B}
satellite scheduled for launch in 2005 August should prove ideal
for this kind of observation. 

\acknowledgments This research was supported by ONR/NRL 6.1 basic
research funds.  I thank Erwin Verwichte for providing the
\textit{TRACE} data for
Figure~\ref{fig:bcs_sumer_period_vs_decay} in machine readable
form, and H.~P. Warren, G.~A. Doschek, and the anonymous referee
for their helpful comments on the manuscript. 

\bibliographystyle{apj}
\bibliography{allrefs}

\clearpage

\begin{deluxetable}{llllllll}
\tabletypesize{\scriptsize}
\tablecolumns{8}
\tablewidth{0pc}
\tablecaption{Doppler Shift Oscillation Properties}
\tablehead{
\colhead{} & 
\colhead{} &
\colhead{} & 
\colhead{} &
\colhead{$A_0$} & 
\colhead{$\omega$} &
\colhead{$\phi$} & 
\colhead{$\lambda$} \\
\multicolumn{2}{c}{$t_0$ (UT)} &
\colhead{Class} & 
\colhead{Location} & 
\colhead{(km s$^{-1}$)} & 
\colhead{(rad min$^{-1}$)} & 
\colhead{(rad)} & 
\colhead{(min$^{-1}$)}
 }
\startdata
1991 Oct  7 & 10:16:57 & C9.9 & S13W82 & 
$66.18 \pm  5.91$ & $1.36 \pm  0.04$ & $-5.80 \pm 0.07$ & $0.273 \pm 0.047$ \\
  &   &   &   & 
$38.92 \pm  11.54$ & $1.27 \pm  0.10$ & $-5.63 \pm 0.06$ & $0.653 \pm 0.138$ \\
1991 Oct 21 & 02:47:34 & C4.3 & S06E89y & 
$18.89 \pm  1.04$ & $0.99 \pm  0.04$ & $-5.35 \pm 0.17$ & $0.276 \pm 0.020$ \\
  &   &   &   & 
$10.82 \pm  2.22$ & $1.17 \pm  0.12$ & $-5.54 \pm 0.17$ & $0.497 \pm 0.131$ \\
1992 Jan 15 & 08:59:21 & C5.0 & S18W90 & 
$6.93 \pm  0.76$ & $1.39 \pm  0.09$ & $-5.54 \pm 0.17$ & $0.388 \pm 0.086$ \\
  &   &   &   & 
$9.69 \pm  2.10$ & $1.10 \pm  0.08$ & $-4.98 \pm 0.30$ & $0.253 \pm 0.068$ \\
1992 Feb 26 & 01:35:57 & M1.3 & S15W90 & 
$22.63 \pm  2.74$ & $2.47 \pm  0.11$ & $-5.82 \pm 0.21$ & $0.498 \pm 0.056$ \\
  &   &   &   & 
$20.40 \pm  6.11$ & $1.93 \pm  0.10$ & $-4.00 \pm 0.27$ & $0.298 \pm 0.115$ \\
1992 Mar 16 & 06:06:15 & C5.8 & S25W21 & 
$9.09 \pm  0.28$ & $1.32 \pm  0.05$ & $-0.84 \pm 0.04$ & $0.404 \pm 0.028$ \\
  &   &   &   & 
$7.93 \pm  2.17$ & $1.44 \pm  0.07$ & $-0.24 \pm 0.20$ & $0.160 \pm 0.110$ \\
1992 Apr 19 & 02:11:32 & C3.9 & N15E78 & 
$6.49 \pm  1.74$ & $2.16 \pm  0.05$ & $-0.50 \pm 0.20$ & $0.117 \pm 0.063$ \\
  &   &   &   & 
$4.38 \pm  2.96$ & $1.58 \pm  0.23$ & $-5.17 \pm 0.59$ & $0.193 \pm 0.285$ \\
1992 Apr 20 & 06:33:25 & C3.6 & S15E39 & 
$5.89 \pm  1.05$ & $1.94 \pm  0.05$ & $-4.69 \pm 0.21$ & $0.149 \pm 0.043$ \\
& & & & \nodata & \nodata & \nodata & \nodata \\
1992 May 27 & 14:37:53 & C7.1 & N23W59 & 
$11.01 \pm  2.05$ & $2.48 \pm  0.06$ & $-3.58 \pm 0.13$ & $0.391 \pm 0.090$ \\
  &   &   &   & 
$6.00 \pm  1.84$ & $2.22 \pm  0.07$ & $-2.30 \pm 0.26$ & $0.117 \pm 0.075$ \\
1992 Jul  4 & 22:49:16 & C6.9 & S12E89 & 
$6.22 \pm  1.93$ & $0.67 \pm  0.06$ & $-2.94 \pm 0.13$ & $0.229 \pm 0.078$ \\
  &   &   &   & 
$6.45 \pm  1.10$ & $0.58 \pm  0.08$ & $-2.74 \pm 0.22$ & $0.181 \pm 0.098$ \\
1992 Jul 17 & 22:36:31 & C5.3 & S11W87 & 
$4.49 \pm  1.23$ & $2.09 \pm  0.05$ & $-2.62 \pm 0.17$ & $0.214 \pm 0.079$ \\
  &   &   &   & 
$7.87 \pm  4.82$ & $1.76 \pm  0.11$ & $-1.16 \pm 0.52$ & $0.202 \pm 0.139$ \\
1992 Aug 24 & 01:12:48 & C2.2 & N14W90 & 
\nodata & \nodata & \nodata & \nodata \\ & & & &
$7.58 \pm  2.94$ & $2.99 \pm  0.25$ & $-2.90 \pm 0.40$ & $0.474 \pm 0.217$ \\
1992 Oct 12 & 21:50:44 & C2.5 & S16W88 & 
$11.37 \pm  1.87$ & $1.60 \pm  0.06$ & $-2.43 \pm 0.05$ & $0.284 \pm 0.090$ \\
  &   &   &   & 
$8.77 \pm  2.79$ & $1.76 \pm  0.17$ & $-2.86 \pm 0.41$ & $0.235 \pm 0.130$ \\
1992 Oct 27 & 22:14:55 & C5.4 & N08W90 & 
$14.03 \pm  1.44$ & $0.96 \pm  0.04$ & $-1.07 \pm 0.20$ & $0.165 \pm 0.028$ \\
  &   &   &   & 
$16.96 \pm  3.94$ & $1.25 \pm  0.14$ & $-2.18 \pm 0.46$ & $0.156 \pm 0.107$ \\
1992 Nov  5 & 06:24:08 & M2.0 & S17W88 & 
$9.22 \pm  0.60$ & $0.92 \pm  0.02$ & $-4.37 \pm 0.11$ & $0.072 \pm 0.017$ \\
  &   &   &   & 
$24.41 \pm  2.00$ & $0.74 \pm  0.02$ & $-2.90 \pm 0.10$ & $0.159 \pm 0.018$ \\
1993 Feb  1 & 10:07:38 & C6.9 & S11E90 & 
$3.65 \pm  0.32$ & $1.55 \pm  0.23$ & $-1.94 \pm 0.54$ & $0.270 \pm 0.058$ \\
  &   &   &   & 
$3.52 \pm  1.66$ & $1.69 \pm  0.26$ & $-1.64 \pm 0.55$ & $0.253 \pm 0.219$ \\
1993 Apr 15 & 09:08:36 & C1.2 & S19W90y & 
$26.00 \pm  1.52$ & $0.78 \pm  0.01$ & $-1.60 \pm 0.05$ & $0.142 \pm 0.009$ \\
  &   &   &   & 
$42.19 \pm  4.08$ & $0.83 \pm  0.03$ & $-2.04 \pm 0.14$ & $0.193 \pm 0.016$ \\
1993 Sep 26 & 10:22:47 & C3.4 & N12E90 & 
$15.43 \pm  4.23$ & $0.96 \pm  0.03$ & $-3.69 \pm 0.25$ & $0.173 \pm 0.030$ \\
  &   &   &   & 
$76.32 \pm  17.88$ & $0.42 \pm  0.03$ & $-0.68 \pm 0.04$ & $0.193 \pm 0.053$ \\
1994 Feb 27 & 09:11:09 & M2.8 & N08W88 & 
$3.60 \pm  0.60$ & $1.39 \pm  0.03$ & $-0.55 \pm 0.04$ & $0.143 \pm 0.031$ \\
  &   &   &   & 
$4.75 \pm  1.16$ & $1.40 \pm  0.04$ & $-0.67 \pm 0.19$ & $0.190 \pm 0.055$ \\
1999 Oct 25 & 06:36:18 & M1.7 & S18E88y & 
$17.25 \pm  1.52$ & $0.65 \pm  0.02$ & $-0.10 \pm 0.09$ & $0.100 \pm 0.017$ \\
  &   &   &   & 
$33.17 \pm  4.29$ & $0.69 \pm  0.02$ & $-0.35 \pm 0.06$ & $0.178 \pm 0.037$ \\
2000 Jul 14 & 00:41:45 & M1.5 & N20W80y & 
\nodata & \nodata & \nodata & \nodata \\ & & & &
$43.01 \pm  7.50$ & $0.96 \pm  0.04$ & $-5.88 \pm 0.12$ & $0.341 \pm 0.057$ \\
\enddata
\label{table:properties}
\end{deluxetable}

\begin{deluxetable}{llllll}
\tabletypesize{\footnotesize}
\tablewidth{0pc}
\tablecaption{Average Doppler Shift Oscillation Properties}
\tablehead{
\colhead{Parameter} & 
\colhead{\ion{S}{15}} &
\colhead{\ion{Ca}{19}} & 
\colhead{Combined} &
\colhead{\textit{TRACE}} &
\colhead{SUMER}
 }
\startdata
Oscillation Period (min) & $5.2 \pm 2.2$  & $5.8 \pm 3.2$ & 
$5.5 \pm 2.7$ & $5.4 \pm 2.3$ & $17.6 \pm 5.4$ \\
Decay Time (min) & $5.5 \pm 3.1$ & $4.7 \pm 1.8$ & 
$5.0 \pm 2.5$ & $9.7 \pm 6.4$ & $14.6 \pm 7.0$ \\
Amplitude (km s$^{-1}$) & $14.4 \pm 14.5$ & 
$19.6 \pm 19.2$ & $17.1 \pm 17.0$ &
$42 \pm 53$ & $98 \pm 75$ \\
Displacement (km) & $730 \pm 742$ & $1400 \pm 2260$ & 
$1070 \pm 1710$ & $2200 \pm 2800$ & $12500 \pm 9900$ \\
Decay Time to Period Ratio & $1.13 \pm 0.65$ & $0.98 \pm 0.62$ &
$1.05 \pm 0.63$ & $1.8$ & $0.85 \pm 0.35$
\enddata 
\label{table:average_values}
\end{deluxetable}

\begin{deluxetable}{lcc}
\tablecaption{Intensity Fitting Results\label{tab:intens_results}}
\tablewidth{0pt}
\tablehead{
\colhead{Parameter} & \colhead{\ion{S}{15}} &
\colhead{\ion{Ca}{19}}}
\startdata
$A_0$ (counts s$^{-1}$) & $41.5 \pm 8.0$ & $38.8 \pm 7.1$ \\
$\omega$ (rad min$^{-1}$) & $1.05 \pm 0.04$ & $0.81 \pm 0.04$ \\
$\phi$ (rad) & $-0.72 \pm 0.25$ & $0.53 \pm 0.25$ \\
$\lambda$ (min$^{-1}$) & $0.12 \pm 0.04$ & $ 0.14 \pm 0.02$ \\
\enddata
\end{deluxetable}

\end{document}